\documentclass[%
preprint,
 amsmath,amssymb,
 aps,
pra,
]{revtex4-2}

\usepackage{xcolor}
\usepackage{subfigure}
\usepackage{graphicx}
\usepackage{dcolumn}
\usepackage{bm}
\usepackage{hyperref}
\usepackage{epstopdf}

\begin{document}

\title{Direct Electrical Detection of Spin Chemical Potential Due to Spin Hall Effect in $\beta$-Tungsten and Platinum Using a Pair of Ferromagnetic and Normal Metal Voltage Probes}

\author{Soumik Aon} 
\author{Abu Bakkar Miah}
\author{Arpita Mandal} 
\author{Harekrishna Bhunia} 
\author{Dhananjaya Mahapatra} 
\author{Partha Mitra} 
\email{pmitra@iiserkol.ac.in}
 
\affiliation{
 Department of Physical Sciences, Indian Institute of Science Education and Research, Kolkata, Mohanpur 741246, India 
}

\begin{abstract}
The phenomenon of Spin Hall Effect (SHE) generates a pure spin current transverse to an applied current in materials with strong spin-orbit coupling, although not detectable through conventional electrical measurement. An intuitive Hall effect like measurement configuration is implemented to directly measure pure spin chemical potential of the accumulated spins at the edges of heavy metal (HM) channels that generates large SHE. A pair of transverse linearly aligned voltage probes in placed in ohmic contact with the top surface of HM , one being a ferromagnetic metal (FM) with non-zero spin polarization and other is the reference metal (RM) with zero polarization of carriers. This combination of FM/RM electrodes is shown to induce an additional voltage proportional to a spin accumulation potential, which is anti symmetric with respect to opposite orientations of FM controlled by a 2D vector magnet.  Proof of concept of the measurement scheme is verified by comparing the signs of voltages for HM channels of Tungsten (W) and Platinum (Pt) which are known to generate opposite spin accumulation under similar conditions of applied current.  The same devices are also  able to detect the reciprocal effect, inverse spin Hall effect (ISHE) by swapping the current and voltage leads and the results are consistent with reciprocity principle. Further, exploiting a characteristic  feature of W thin film deposition,  a series of devices were fabricated with  W resistivity varying over a wide range of 10 - 750 $\mu \Omega$-cm  and the calculated spin Hall resistivity exhibits a pronounced power law dependence on resistivity. Our measurement scheme combined with almost two decades of HM resistivity variation provides the ideal platform required to test the underlying microscopic mechanism responsible for SHE/ISHE.
\end{abstract} 
\maketitle
   \section{\label{sec:level1}Introduction}
Spin Hall effect (SHE) \cite{hirsch,zhang,sinova2015spin} has emerged as a prominent  transport phenomenon arising explicitly due to the spin degree of the carriers in conductors and relativistic effects. Typically, materials with large  spin-orbit coupling (SOC) strength, e.g. semiconductors, \cite{Kato_2004} heavy metals (HM), \cite{morota2011,pai2012spin} and topological insulators \cite{Gao_2011} etc. exhibits  measurable SHE that results in the conversion of a charge current density ($\textbf{J}_{c}$) into a transverse pure spin current density ($\textbf{J}_{s}$) of spin polarization direction $\textbf{s}$, described by the phenomenological equation. 
\begin{equation}\label{eq:1} 
\textbf{J}_s = \theta_{SHE} (\frac{\hbar}{2e}) \textbf{J}_c \times \textbf{s}, 
\end{equation}
where, the factor $\hbar/2e$ makes the dimensions of $\textbf{J}_s$ same as that of $\textbf{J}_c$.
$\theta_{SHE}$ is the spin Hall angle, a dimensionless material specific parameter whose magnitude quantifies the efficiency of charge-to-spin current conversion and sign determines the orientation of deflected spins ($+\textbf{s}$ or$-\textbf{s}$) in a particular transverse direction.  In a conventional Hall effect like geometry with open circuit condition transverse to current, SHE will lead to accumulation of opposite spins \cite{zhang} at the opposite edges of a current carrying conductor. For normal (paramagnetic) metals with zero spin polarization in the conduction band, a pure spin current is generated by SHE implying that equal number of oppositely oriented spins move in opposite directions so that no net charge current exists. This will lead to opposite spin accumulations at the edges of equal  magnitude and hence no net electric potential difference. Unlike a charge current, a spin current is not a conserved quantity and decays rapidly while diffusing through the conductor within a characteristic length known as the spin diffusion length $\lambda_{sf}$, typically few nano-meters at room temperature. Thus, only the carriers within $\lambda_{sf}$ distance from the transverse edges of the conducting channel will contribute to the spin accumulation.  A rigorous expression for spin potential difference at the edges was derived and in the limit of conducting channel width being much larger than $\lambda_{sf}$, the maximum limit of the spin potential difference achieved is shown to be, \cite{zhang} 
\begin{equation}\label{eq:spinacc} 
\mu_{\uparrow}-\mu_{\downarrow}=\lambda_{sf}\rho_{SHE}J_c
\end{equation}
Where, $\rho_{SHE}$ is the transverse resistivity due to SHE. The reciprocal phenomenon of SHE is the inverse spin Hall effect (ISHE) where a transverse charge current is generated in response to a spin current present in the paramagnetic conductors with strong SOC, described as
\begin{equation}\label{eq:2} 
\textbf{J}_{c} = \theta_{SHE} (-\frac{2e}{\hbar})\textbf{J}_{s} \times \textbf{s}.
\end{equation}

\begin{figure*}
\centering
\includegraphics[width=0.9\textwidth]{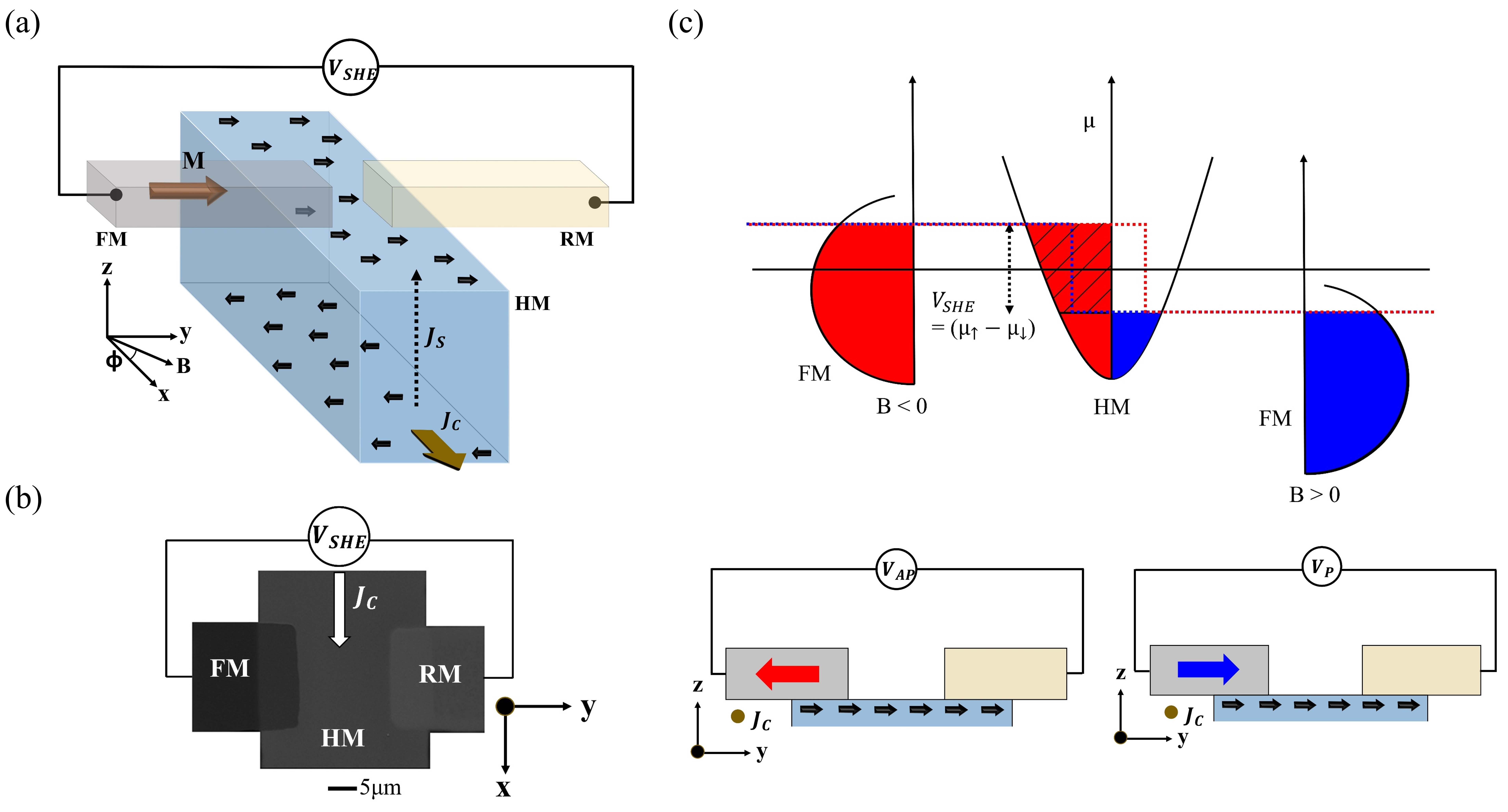}
\caption{(a) The schematic diagram of a device for detection of SHE. A rectangular HM channel (light blue) with long axis chosen to be $x$- axis is embedded in an insulating $SiO_2$ substrate (not shown). A pair of transverse (along $y$) linearly aligned FM (light grey) and RM (light golden) electrodes are in Ohmic contact with on top surface of HM (light blue). Due to SHE a pure spin current $J_s$ flows along $z$, the entire top (bottom) surface of HM assumes spin accumulation with polarisation along $+(-)y$  shown by small black arrows. The device is mounted inside a 2D vector electromagnet (not shown) such that the external magnetic field ($B$) could be varied in the $xy$-plane making an angle $\phi$ in general with the HM channel. The magnetization of the FM (deep orange arrow) is shown here to be along the $y$-direction $\phi=90^0$ when the effect explained in (c) is maximum, (b) FESEM image of a typical device (top view), (c) Illustration of the process of development of excess voltage in FM due to spin accumulation on top surface of HM. The band diagram of HM top surface showing non equilibrium spin accumulation  of $+y$ polarization shown as hatched red region (exaggerated) on an otherwise parabolic band of normal metal with equal occupation of both spin band (blue (red) for $+(-)y$).  An ideal half metallic FM with shallow d-band of one spin orientation is shown for simplicity of illustration. The transverse FM and RM electrodes, connected to voltmeter creates an open circuit condition. As the FM magnetization is flipped with the application of $B$, the Fermi level of FM needs to align with the corresponding spin band of the HM which is unaffected by $B$, thus resulting in a measurable voltage difference proportional to spin accumulation. The side view of the interface between FM and RM with HM are shown in the bottom for opposite FM orientations.}
\label{Device_schematic_SHE}
\end{figure*}

The spin dependent transverse deflections are responsible for generating ISHE is same as that of SHE. However as the incident current is spin polarized, unequal number of carriers are accumulated at the opposite edges, resulting in a finite voltage. In ferromagnetic metals (FM), strong exchange interaction naturally gives rise to a band structure with  finite spin polarization of carriers.  Thus the above mechanisms arising from SOC applied to FM leads to a measurable transverse voltage, which is the phenomenon of Anomalous Hall effect (AHE), a topic of extensive research spanning over several decades. \cite{nagaosa2010anomalous} The main point of contention was to establish whether the underlying spin dependent transverse deflections can arises from scattering events (extrinsic mechanisms) \cite{Smit_1958,Berger_1970,crepieux2001theory} or from band structures effects due to non-zero Berry phase (intrinsic mechanisms). \cite{karplus1954hall,Schulz_1990,Sinova_2004} A scaling relation between transverse and longitudinal resistivity of the form $\rho_{xy} \sim \rho_{xx}^n$ is found to hold true in general and the values of the exponent $n$ were associated with specific mechanisms\cite{nagaosa2010anomalous}.  The generally accepted view is that all mechanisms could in principle contribute to AHE, but depending on nature of the material or on the degree of disorder in a given material one mechanism may be dominant over others.  Since the underlying mechanisms behind SHE (or ISHE) is same as that of AHE, the same concepts can be put to test, provided one can overcome  the experimental challenge posed by the lack of a straightforward method to directly quantify the presence of spin chemical potential gradient. In this report we propose a simple intuitive device structure and measurement scheme that can directly measure the spin chemical potential of the accumulated spins due to SHE and calculate $\rho_{SHE}$. Tungsten (W) is a heavy atomic weight metal (HM) a, with highest reported negative $\theta_{SHE}$,  with an useful feature of tunability of crystal structure and morphology by controlling growth conditions, which in turn can lead to large variation of resistivity. Thus we have chosen W as the conducting medium for observing SHE and fabricated series of devices by varying only the resistivity of W ($\rho_W$), keeping other parameters same, so that the scaling behavior of $\rho_{SHE}$ with $\rho_{W}$ could be examined. \\
\begin{figure}
\centering
\includegraphics[width=1.0\textwidth]{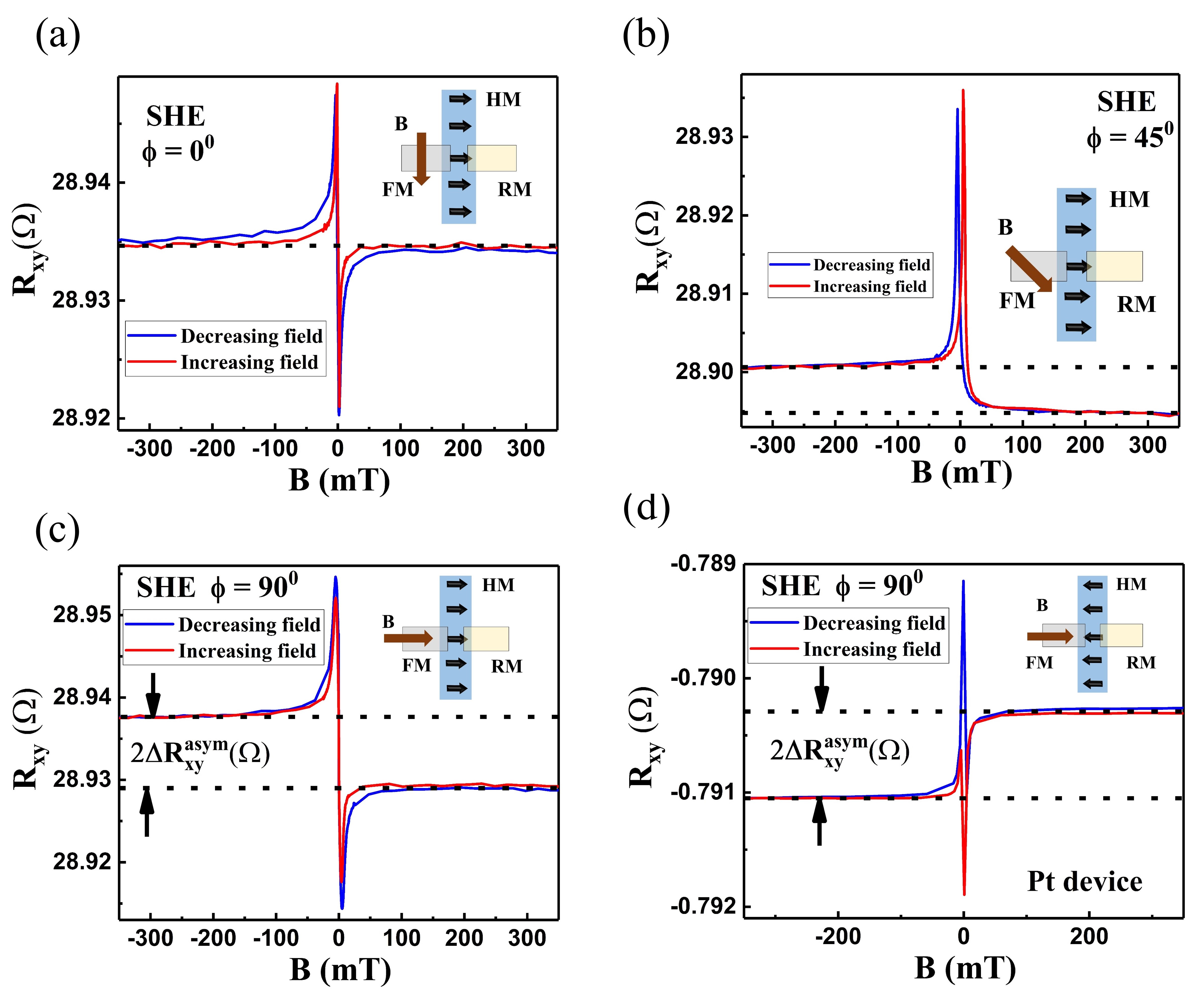}
\caption{Hysteresis response of $R_{xy}(B)$ (=$V_{xy}/I$) in the SHE configuration for $\phi$ = (a) $0^{0}$, (b) $45^{0}$, and (c) $90^{0}$ for a pure $\beta$-W device with $\rho_{W}$ = 750 $\mu\Omega$-cm and (d) for a Pt device for $\phi=90^0$. Insets show the top view of the device schematic showing orientation of the FM magnetization (deep orange arrow)  with respect to fixed spin accumulation (black arrow) in the HM top surface. The dashed lines are eye estimation of approximate saturation values of $R_{xy}$ at high positive and negative fields for a given $\phi$. The difference in the saturation values is denoted as $\Delta R_{xy}^{asym}$ which is found to be (a) zero for $\phi=0^{0}$, (c) maximum for $\phi=90^{0}$ and (b) intermediate for $\phi=45^{0}$ and (d) opposite sign for Pt device at $\phi=90^{0}$ compared to (c).}
\label{Linear_SHE}
\end{figure}

\subsection{\label{sec:level2}Results and Discussion}
An elegant and definitive method of detecting spin chemical potential was demonstrated in the context of lateral spin valve in the so called non-local measurement geometry \cite{Johnson_1985,johnson1988coupling} where pure spin current was injected into a normal metal from one FM electrode and a voltage proportional to spin accumulation was detected by another FM electrode placed within spin diffusion length of the spin injector. Subsequently, in the seminal work on the theoretical framework for SHE in diffusive limit, a measurement scheme was proposed  using a FM voltage probe \cite{zhang} attached to one of transverse edges in current carrying conductor. The finite spin polarization ($P$) of the FM conduction band was  crucial for such measurements. Early experimental works on electrical detection of SHE was primarily based on converting the generated spin current into an electric voltage due to ISHE using complex device structures\cite{valenzuela2006direct,kimura,niimi2011extrinsic}. An important step towards direct electrical detection of SHE was reported \cite{pham-2016} using a device featuring two FM electrodes (CoFe) of different coercive fields placed on a HM channel (Pt), and the voltage developed between the two FM electrodes were demonstrated to vary with hysteretically  between parallel and anti-parallel states. Under the assumption that the two FM electrodes are exactly same, for the parallel states the measured voltage is supposed to give zero SHE contribution , while for anti-parallel sates the effect of SHE adds up , and hence the half of the width of the transverse voltage  hysteresis loop was identified as the spin Hall voltage.\\ 

\begin{figure}
\centering
\includegraphics[width=1.0\textwidth]{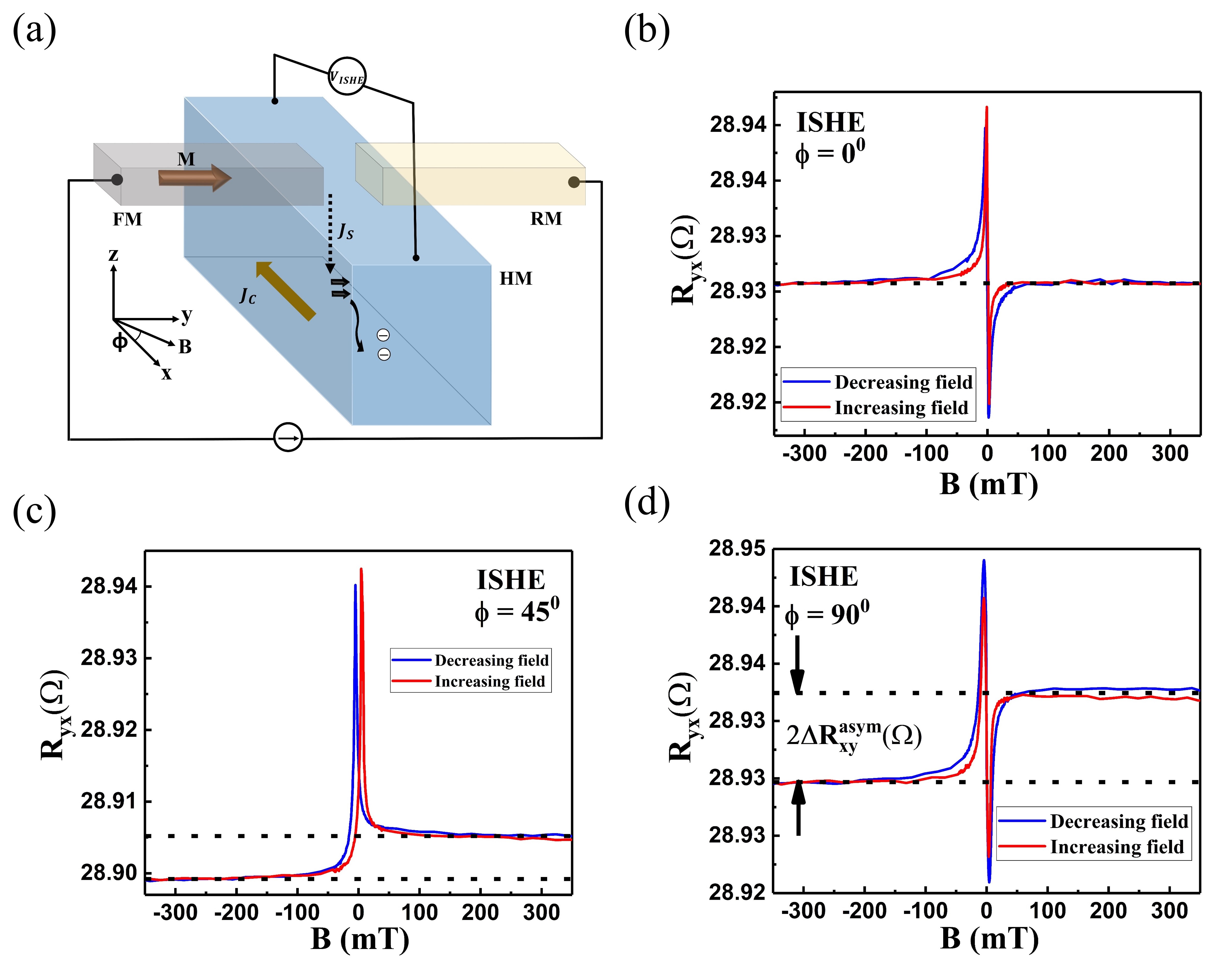}
\caption{(a) The schematic of the device in ISHE configuration. The current source is connected between FM and RM and voltage is measured between the ends of the HM channel. The current at the FM/HM interface will inject a spin current ($J_s$) in the HM along the $-z$ direction. Assuming an ideal single spin band ferromagnetic for simplicity, it is shown that due to ISHE all the spins are deflected in the $+x$ direction, developing a measurable voltage in the HM. The hysteresis behaviour of the transverse resistance $R_{yx}(=V_{yx}/I)$ in the ISHE configuration for the $\beta$-W device with $\rho_{W}$ = 750 $\mu\Omega$-cm for angles $\phi$ = (b) $0^{0}$, (c) $45^{0}$, and (d) $90^{0}$. The dashed lines are for eye estimation of approximate saturation values of $R_{yx}$ at high positive and negative fields for a given $\phi$. The difference in the saturation values is denoted as $\Delta R_{yx}^{asym}$ which is found to be (b) zero for $\phi=0^{0}$, (d) maximum for $\phi=90^{0}$ and (b) intermediate for $\phi=45^{0}$.}
\label{Linear_scan_ISHE}
\end{figure}
We propose a device structure where a transverse voltage in response to a current in the HM channel is recorded using a combination of one FM electrode along with a paramagnetic metal electrode, denoted as reference metal (RM) as shown schematically in Figure \ref{Device_schematic_SHE}(a) and explained in the figure caption. A FESEM image of a typical device is shown in Figure \ref{Device_schematic_SHE}(b). Details about the device fabrication steps can be found in the Materials and Methods section in Supplementary Information. We highlight a few important refinements incorporated in our methods. Firstly, steps were taken to ensure that the HM channel is embedded into insulating the $SiO_2$ substrate with only top surface exposed so that the FM/RM electrodes are in contact only with the spin accumulation at the top surface of HM. Secondly, we use a 2D vector magnet and mount the substrate appropriately so that the magnetic field can rotate the FM magnetization in the plane of the device, while the spin accumulation due to SHE remains unaffected. The in plane orientation of the magnetic field vector ($\textbf{B}$) is determined by the azimuthal angle $\phi$ as shown in Figure \ref{Device_schematic_SHE} (a-b). The underlying principle for the direct measurement of spin potential is explained in Figure \ref{Device_schematic_SHE}(c) for $\phi=90^0$ when the effect is maximum. For HM materials with positive $\theta_{SHE}$, equation \ref{eq:1} implies that a current in the HM channel (along $x$) results in a transverse spin current (along $z$) and the entire top (bottom) surface of the HM assumes spin accumulation in positive (negative) $y$ direction. The non-equilibrium spin accumulation on the top surface (along $+y$) is depicted as the hatched red region (exaggerated for clarity) of the HM band diagram. For simplicity of understanding, we consider the FM electrode to be a  half metal (single spin sub-band at the Fermi level, P=1). Depending on whether the FM magnetization is parallel or anti-parallel to the accumulated spin on the top surface of HM, the Fermi level of the FM will align with the chemical potential of either $+y$ or $-y$ spin band of HM. In contrast, the Fermi level of a normal metal reference electrode (RM) electrode (P=0)  will align with the average of the two spin chemical potentials of the HM under all conditions. Thus, a voltmeter connected between a pair of FM and RM electrode in contact with the HM, will register a voltage difference when magnetization of FM is reversed (with the help of an external magnetic field) and can be shown to be proportional to the difference in spin potentials. A rigorous derivation for a general FM ($P<1$) shows the voltage \cite{zhang} to be read in conjunction with equation \ref{eq:spinacc}.
\begin{equation}\label{deltaV}
   \Delta V=P(\mu_{\uparrow}-\mu_{\downarrow}) 
\end{equation}

A proof of concept of the above idea lies in the fact that the sign of $\Delta V$ deduced above will reverse for HM material with negative $\theta_{SHE}$ and to test this idea, devices were fabricated with Pt (positive $\theta_{SHE}$) \cite{Ando_2008,Lui_SHE_2011} and W (negative $\theta_{SHE}$) \cite{pai2012spin,hao2015giant,hao2015beta} as HM material. The magnetic field dependence of $R_{xy}= V_{xy}/I$ for the highest resistivity $\beta$-W device ($\rho_{W}$ = 750 $\mu\Omega$-cm) at three representative angles of $\phi$ = $0^{0}$, $45^{0}$, and $90^{0}$ are shown in Figure \ref{Linear_SHE}(a)-(c). For $\phi=0^{0}$ [Figure \ref{Linear_SHE}(a)], corresponding to the configuration when FM magnetization is always orthogonal to spin accumulation (see inset) and as expected there is no change in $R_{xy}$ when magnetization is flipped. In contrast, for $\phi=90^{0}$ [Figure \ref{Linear_SHE}(c)], biggest change in $R_{xy}$ is observed as the magnetization is flipped  from being parallel to anti-parallel to the spin accumulation. For $\phi$ = $45^{0}$ configuration [Figure \ref{Linear_SHE}(b)], a component of total magnetization of FM, changes from being parallel to anti-parallel to the spin accumulation as the magnetization is flipped, resulting in an intermediate change in $R_{xy}$. Alternatively, the Pt-based device depicted in \ref{Linear_SHE}(d) exhibits a opposite field dependent transverse resistance as compared to the W device at $\phi$=$90^{0}$. This contrast could be attributed to the voltage induced by opposite spin accumulation (compared to W) generated at the top and bottom surface of the device. The same devices can also be used for detecting ISHE in the HM by swapping the current and voltage leads as shown schematically in Figure \ref{Linear_scan_ISHE}(a). A spin polarized current is injected from the FM into the HM along $-z$ direction in the overlap region with spin polarization $\bf{s}$ along $+y$. The SOC driven deflection of injected spins will result in a transverse charge current [equation \ref{eq:2}] towards $-x$ direction along the HM channel which will thus register a finite transverse voltage across its ends. Once the magnetization of FM is flipped, the polarization of the injected spins and hence the sign of voltage is also reversed. The magnetic hysteresis curves of the transverse resistance in the ISHE configuration, defined as $R_{yx}=V_{yx}/I$ for the same device of $\rho_{W}$ = 750 $\mu\Omega$-cm as shown in Figure \ref{Linear_scan_ISHE}(b)-(d). For $\phi$=$0^{0}$, the FM is always magnetized along the HM channel and thus there is no charge current and hence voltage along the HM channel. Thus we observe that $R_{yx}$ values for opposite magnetization orientation is same [Figure \ref{Linear_scan_ISHE}(b)]. In contrast, for $\phi$=$90^{0}$ the ISHE charge current generated along the HM channel and the change in the $R_{yx}$ upon flipping of FM is maximum [Figure \ref{Linear_scan_ISHE}(d)]. For $\phi$=$45^{0}$ the change in $R_{yx}$ is intermediate as a component of total magnetization contributes to a transverse voltage along the HM channel [Figure \ref{Linear_scan_ISHE}(c)]. The measurements of SHE and ISHE hysteresis scans on each devices at various angles seems to conform with the Onsager reciprocity principle.\\
\begin{figure}
\centering
\includegraphics[width=0.8\textwidth]{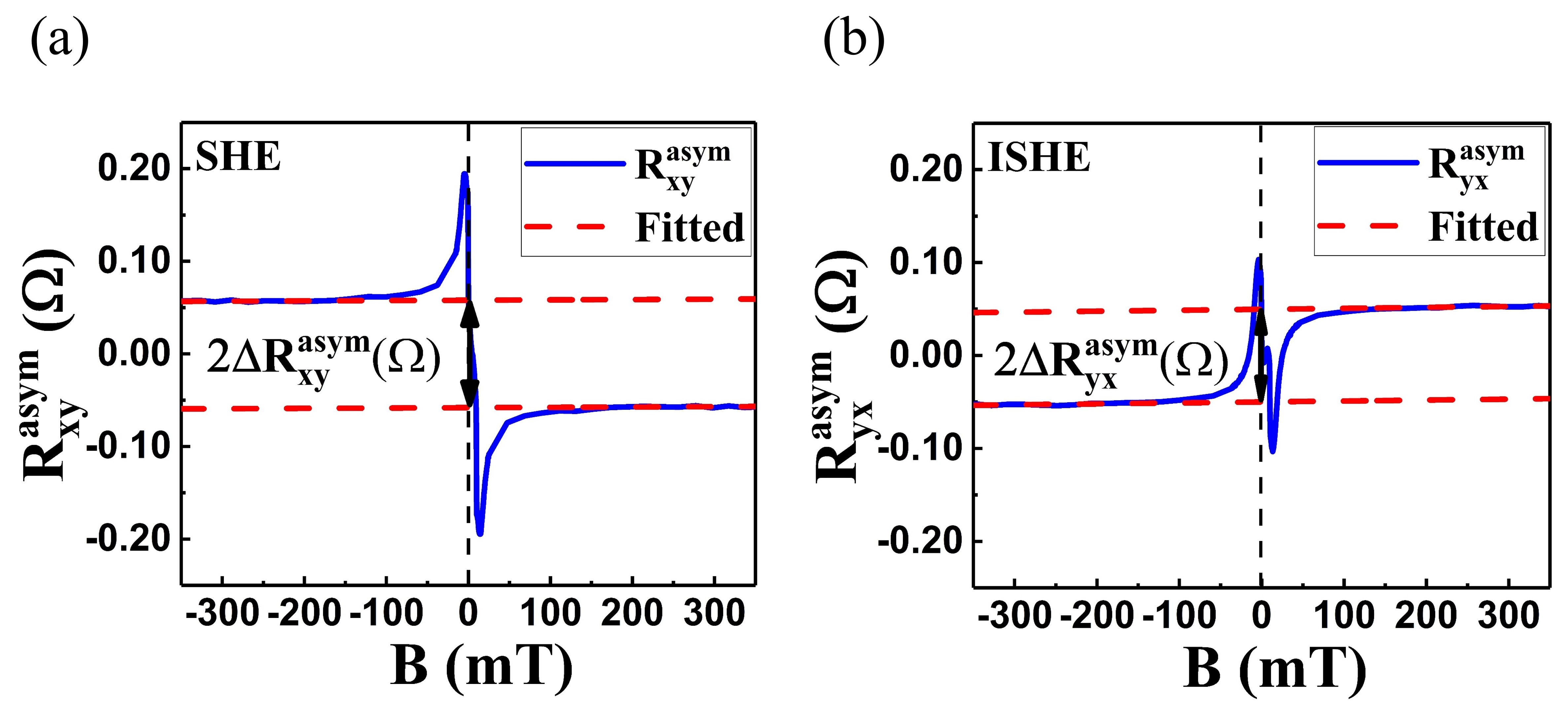}
\caption{Plot of the asymmetric contribution of the as measured transverse resistance in (a) SHE configuration $R^{asym}_{xy}(B)$ for data shown in Figure \ref{Linear_SHE}(c) and ISHE configuration $R^{asym}_{yx}(B)$ for data shown in Figure \ref{Linear_scan_ISHE}(d). Both plots have same scale on both axis. The data for  $|B|\geq$ 150 mT are linearly fitted  and shown as dashed lines to take into account possible artifacts like ordinary Hall effect or AHE of FM. The intercepts of the linear fits is a measure of the the asymmetric transverse resistance due to SHE at $B=0$ and denoted as  $\Delta R^{asym}_{xy}$ which is almost same in magnitude for SHE and ISHE in agreement with the reciprocity relation of the two phenomenon.}
\label{Asym}
\end{figure} 
The as measured transverse resistance $R_{xy/yx}$ requires three important corrections so that it can be quantitatively correlated with SHE(ISHE) resistivity. Firstly, the raw data presented in Figure \ref{Linear_SHE} for SHE and Figure \ref{Linear_scan_ISHE} for ISHE, is not purely asymmetric with respect to flipping of magnetization of the FM, as demanded by the the proposed measurement models. The presence of a symmetric contribution is typical in any Hall effect measurement, that may arise in general from misalignment of transverse voltage probes, resulting in picking up of longitudinal voltage drop and /or in the current scenario from planner Hall effect \cite{chang_PHE} of the FM. Following standard procedure, the anti-symmetric component of transverse voltage can be extracted as $V_{xy}^{asym}=[V_{xy}(+B)-V_{xy}(-B)]/2$. Secondly, as the HM and FM are in Ohmic contact, an applied current $I$ will redistribute between HM, FM, and RM layers in the overlapped regions. The FM is sensing the addition asymmetric voltage only due to the current flowing through the HM ($I_{HM}$) in contact. Using parallel resistor model and geometry of the devices (details shown in SI), we deduce that. 
\begin{equation}
I_{HM} = \frac{I}{[{{1 + \frac{\rho_{HM}}{t_{HM}w_{HM}}(\frac{t_{FM}w_{FM}}{\rho_{FM}} + \frac{t_{RM}w_{RM}}{\rho_{RM}}}})]},
\end{equation}
where, $\rho_{HM/FM/RM}$, $t_{HM/FM/RM}$ represent the resistivity, thickness of HM, FM, and RM, respectively. $w_{HM}$ refers to the width of HM and $w_{FM/RM}$ are overlapped width of FM and RM layers on HM. A plot of the field dependence of a purely  anti-symmetric transverse resistance defined as $R_{xy}^{asym}=V_{xy}^{asym}/I_{HM}$ is shown in Figure \ref{Asym}. Thirdly, the calculated $R_{xy}^{asym}$ may still have a component arising from ordinary Hall effect (OHE) of the HM channel or AHE of the FM electrode that may arise from slight unavoidable component of $B$ normal to sample plane, both contribution being linear in $B$. To eliminate such artifacts, linear fitting are performed on the high field ($|$B$|$ $>$ 150 mT) regions of $R_{xy}^{asym}(B)$ curves, shown as dashed lines in Figure \ref{Asym}. We identify the extrapolated values of the fitted lines at $B=0$, as the contribution arising solely from SHE or ISHE and is denoted as $\Delta R_{xy}^{asym}$ for a given scan angle $\phi$.\\ 
\begin{figure}[h!]
\centering
\includegraphics[width=0.5\textwidth]{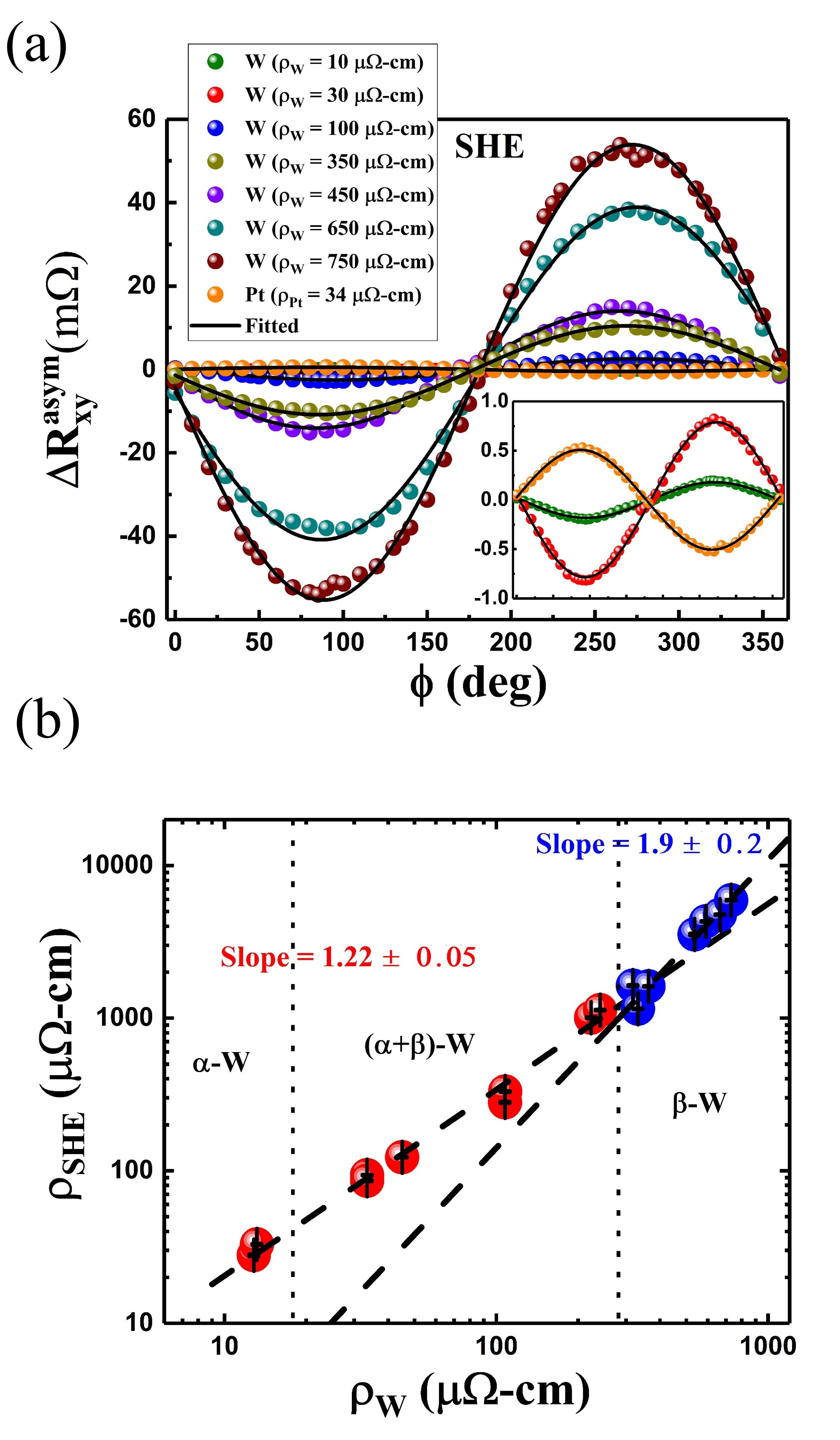}
\caption{(a) Plots of $\Delta R^{asym}_{xy}$ extracted from individual hysteresis scans in SHE configuration for a fixed $\phi$  ranging from $0^{0}$ to $360^{0}$ in steps of $10^{0}$, for series of devices with $\rho_{W}$ = 10, 30, 100, 350, 450, 650, 750 $\mu\Omega$-cm and one Pt device  with $\rho_{Pt}$ = 34 $\mu\Omega$-cm. Inset shows the low resistivity devices plotted separately for clarity. Solid lines are fits to the equation of the form $R_{SHE} sin(\phi-\phi_{0})$, (b) Plot of spin Hall resistivity ($\rho_{SHE}$) for all devices vs. longitudinal resistivity ($\rho_{W}$) in log-log scale. The data points for pure $\beta$-W is shown in blue while that for pure $\alpha$-W and mixed ($\alpha+\beta$)-W  are shown as red. Both sets are  are separately fitted to straight lines and the corresponding slopes representing the power law exponents are indicated.}
\label{Scaling}
\end{figure}
Our experiment extends over a set of devices with different HM channels that includes a  series of W of resistivity $\rho_W$ = 10, 30, 100, 350, 450, 650, 750 $\mu\Omega$-cm, and one Pt of resistivity $\rho_{Pt}$ = 34 $\mu\Omega$-cm, all having same geometric parameters and FM electrode. For each of our devices, we recorded $R_{xy}(B)$ hysteresis scans with in-plane $B$ at fixed $\phi$ ranging from $0^{0}$ to $360^{0}$ in steps of $10^{0}$, in both SHE and ISHE measurement configurations. $\Delta R^{asym}_{xy}$ is calculated for each such scan following the analysis mentioned above. The results are summarised in Figure \ref{Scaling}(a). The inset graph highlights the plots for devices with $\rho_W$ = 10 and 30 $\mu\Omega$-cm and $\rho_{Pt}$ = 34 $\mu\Omega$-cm, as the magnitudes of $\Delta R_{xy}^{asym}$ is comparatively small and distinguishable in the main plots. The variation of $\Delta R_{xy}^{asym}(\phi)$ for all the W devices is exactly opposite to that of the Pt device and is consistent with the opposite signs of $\theta_{SHE}$ of the materials. It may be  that the orientation of the spin accumulation is always transverse ($\pm y$) to the current irrespective of applied field and the orientation of the FM with respect to the channel is determined by $\phi$ so that the component of magnetization along the spin accumulation varies as $sin(\phi)$. We find that for all the devices, variation of $\Delta R_{xy}^{asym}$ appears to fit well with curves of the form $R_{SHE} sin(\phi-\phi_{0})$ [Figure \ref{Scaling}(a)], $R_{SHE}$ signifies the maximum SHE signal when the spin accumulation and FM are parallel and $\phi_{0}$ represents misalignment of the HM channel with the vector magnet $x$-axis, typically $\sim 5^{0}$. Previous reports on measurement of SHE \cite{kimura,groen2021} has confirmed such angular dependence of measured voltage with detector magnetization. The values of $R_{SHE}$ extracted from the fits of the W devices is found to increase with  $\rho_{W}$, with highest value of $\sim$ -55 m$\Omega$ observed in the pure $\beta$-W device of resistivity $\rho_{W}$ = 750 $\mu\Omega$-cm. The $R_{SHE}$ for Pt device is significantly low $\sim$ +0.5 m$\Omega$, which is comparable to the value $\sim$ -0.2 m$\Omega$ observed in pure $\alpha$-W with $\rho_{W}$ = 10 $\mu\Omega$-cm. Exact same analysis performed on the ISHE data for each samples gives $R_{ISHE}$ values [Figure S5 (a) in SI] same as that of corresponding $R_{SHE}$, which is expected from the reciprocity relations.\\

The HM materials in our devices are deposited using magnetron sputtering under high vacuum conditions on amorphous substrates at room temperatures, resulting in disordered polycrystalline films with rough morphology which is confirmed by AFM and SEM images. Hence it is argued that our HM channels are not conducive to support the intrinsic mechanisms  of SHE (ISHE) and only extrinsic mechanisms are dominant. We further argue that resistivity of the HM channels is a measure of disorder \cite{salmon2013structure}, which implies that in a device with higher resistivity, the carriers will undergo more scattering and result in higher transverse spin accumulation in case of SHE configuration or higher charge current in case of ISHE configuration. This qualitatively explains the increase of $R_{SHE}/R_{ISHE}$ with the $\rho_W$ [Figure \ref{Scaling}(a)]. The extrinsic mechanisms are further classified into two broad categories. For the skew scattering mechanism\cite{Smit_1958}, the transverse conductivity was shown to be directly proportional to the mean free path of the material, which translates into dependence of the form $\rho_{xy}\sim\rho_{xx}$. The side jump mechanism \cite{Berger_1970} on the other hand is shown to have a transverse conductivity that is independent of the mean free path which leads to relations of the form $\rho_{xy}\sim\rho_{xx}^2$. We first calculated the transverse resistivity following equation \ref{eq:spinacc} and equation\ref{deltaV} as,
\begin{equation}
    \rho_{SHE}=\frac{R_{SHE} w_{HM} t_{HM} }{P\lambda_{sf}}
\end{equation}
where we have used $P=0.3$ \cite{Zahnd_2016} and $\lambda_{sf}$ = 2 nm \cite{Kim_lambda} for all our devices as we could not independently estimate these parameters. A  plot of $\rho_{SHE}$ as a function of  $\rho_W$ on a log-log scale is shown in Figure \ref{Scaling}(b) and the slopes of linear fits that gives the exponents of the scaling relations $\rho_{xy}\sim \rho_{xx}^{n}$. We have further segregated the data into three regimes of W morphology as indicated by vertical dashed lines: (i) pure metastable $\beta$-W phase (ii) mixed $(\alpha+\beta)$-W  and (iii) pure $\alpha$-W phase. We find, that if the fitting is restricted to pure $\beta$-W samples, the slope obtained is 1.9 $\pm$ 0.2 which is indicative of dominant side jump mechanism. If the linear fit is performed on sample set excluding the pure $\beta-$W devices, the slope obtained is $1.22\pm0.05$, which is indicative of presence of both skew scattering and side-jump mechanism. The same analysis is performed on the ISHE resistivity (shown in SI) and similar trend is observed as expected from reciprocity principle. In the context of AHE, it was established that since transverse conductivity ($\sigma_{xy}$) for side jump mechanism being independent of the mean free path, is dominant for more disordered samples with low mean free path. Our results seems to corroborate the fact in the context of SHE for the highly disordered pure $\beta$-W samples. 

\subsection{Conclusion}
To conclude, we have demonstrated the functioning of an intuitive device configuration that allows for direct detection of spin potential arising out of SHE in HM materials using a combination of FM/HM voltage probes. The same devices were used to detect ISHE and verify the validity of reciprocity relations. The use of a vector magnet to rotate the magnetization of the FM provides a accurate measure of the maximum signal arising from SHE/ISHE and hence resistivity.  Further we have exploited the large tunability of resistivity of W and examined the  scaling relations to identify the underlying microscopic mechanisms and found pronounced evidence for the side jump mechanism in higher resistivity samples. Our measurement scheme can be further generalised to study other spintronics phenomenon. 

\subsection{Acknowledgement}%
We acknowledge the support of IISER Kolkata,  Ministry of Education (MoE), Govt. of India  funding the research and providing fellowship to SA and HB. ABM and DM acknowledge the support of fellowship from CSIR, Govt. of India. PM acknowledges support from grant SR/FTP/PS-077/2011, SERB, Govt. of India.\\

\bibliography{references}

\clearpage  
\onecolumngrid  
\noindent \textbf{\LARGE Supplementary Information for `Direct Electrical Detection of Spin Chemical Potential Due to Spin Hall Effect in $\beta$-Tungsten and Platinum Using a Pair of Ferromagnetic and Normal Metal Voltage Probes'}\\

\setcounter{figure}{0}
\renewcommand{\thefigure}{S\arabic{figure}}

\noindent \textbf{Note 1. Materials and Methods}:\\

\noindent \textbf{Deposition $\&$ Characterization of W-Thin Films of Various Resistivity (HM layer)}:\\

\begin{figure}[h!]    
\centering
\includegraphics[width=1.0\textwidth]{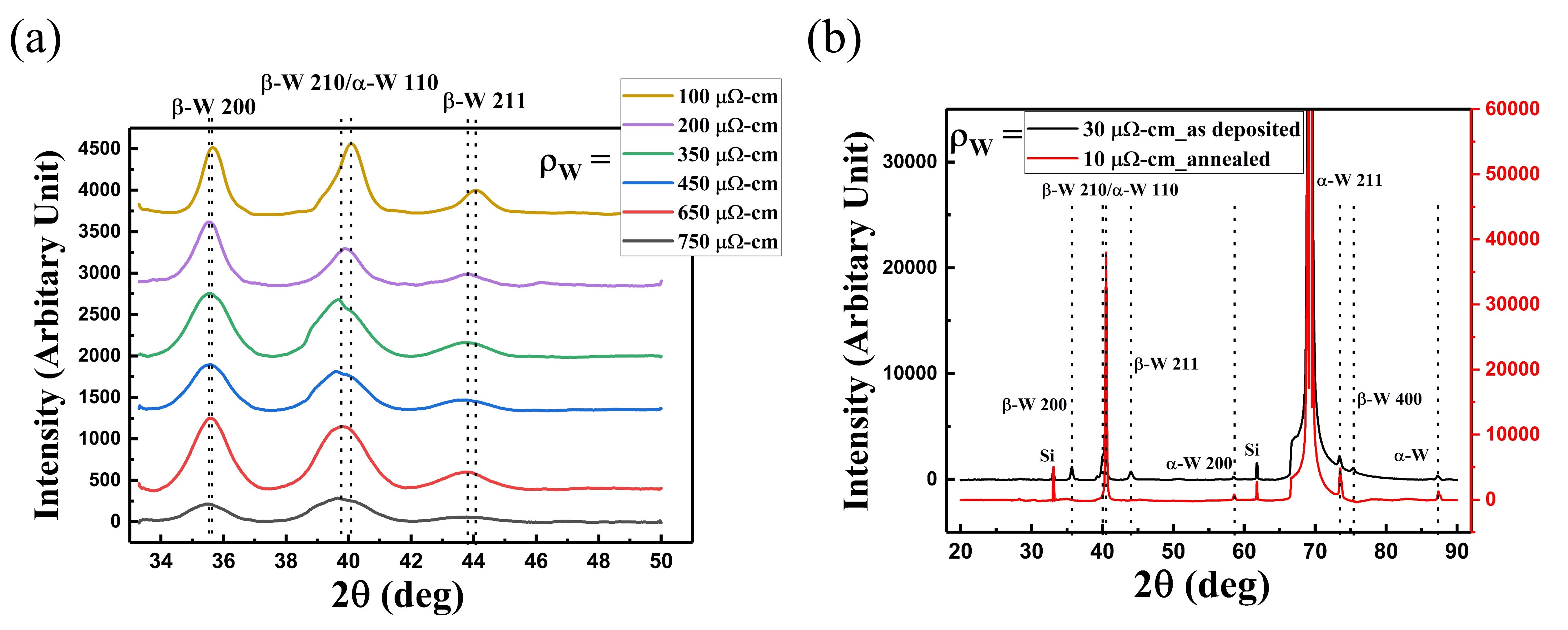}
\caption{XRD patterns of W thin films under varying $P_{Ar}$. The peak positions with their relative peak intensities corresponding to the phases of W are shown as vertical dotted lines. (a) XRD of as-deposited W films with resistivity ranging from 100 $\mu\Omega$-cm to 750 $\mu\Omega$-cm. The films at higher $P_{Ar}$ ($\rho_{W}$ = 750, 650, 450, and 350 $\mu\Omega$-cm) are completely in $\beta$ phase, whereas $\rho_{W}$ = 100, 200 $\mu\Omega$-cm exhibit a dominant $\beta$ phase mixed with some $\alpha$, characterized by the shift in peak position from $2\theta$ = $39.7^{0}$ to $40.04^{0}$. (b) As deposited and annealed W films which are mixed $(\alpha+\beta)$-W and complete $\alpha$-W respectively.}
\label{XRD}
\end{figure}

\begin{figure}[h!]
\centering
\includegraphics[width=1.0\textwidth]{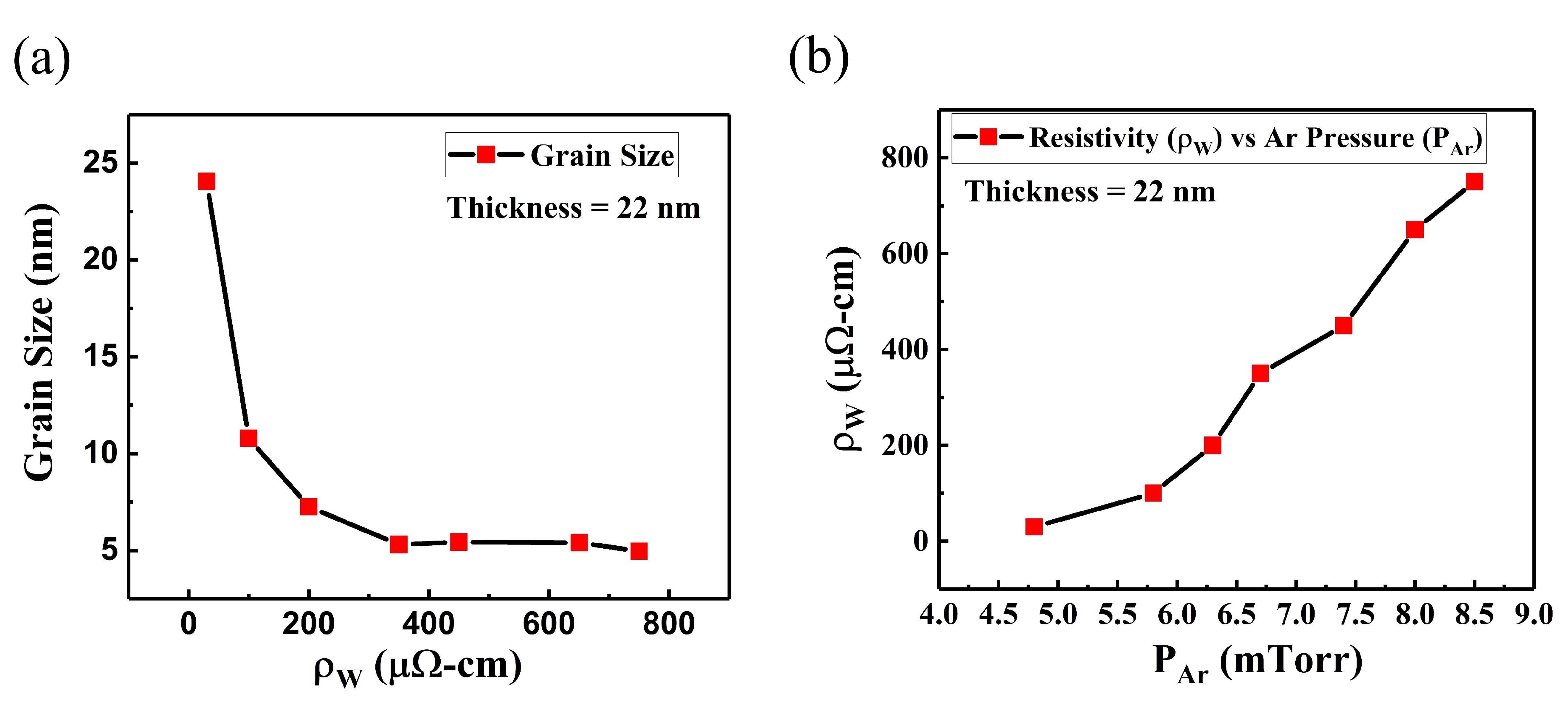}
\caption{(a) depicts how the the grain size changes with $\rho_{W}$. As $\rho_{W}$ decreases, the grain size is observed to increase. The increased magnitude is more pronounced in $\alpha$-W, (b) shows $\rho_{W}$ as a function of $P_{Ar}$. A monotonic increase in $\rho_{W}$ is observed in relation to increasing $P_{Ar}$.}
\label{Grain}
\end{figure}
\noindent 

We prepared a series of 22 nm thick W-thin films on $Si/SiO_{2}$ (300 nm) substrates at room temperature by dc magnetron sputtering (Mantis Deposition LTD) using a combination of scroll and turbo pump achieving a base pressure of $1\times10^{-6}$ torr. The W target (Kurt Lesker), measuring 3 inches in diameter $\times$ 3 mm thickness (99.95\% pure), was maintained at a constant distance of 15 cm from the substrate. A straight deposition setup ensured film homogeneity. These W films were fabricated under varying argon pressures ($P_{Ar}$), while maintaining a constant sputtering power of 100 watt. $P_{Ar}$ was systematically varied, ranging from 4.5 mTorr to 8.5 mTorr. Correspondingly, the deposition rate of W thin films decreased from 1.0 {\AA}/sec for $P_{Ar}$ = 4.5 mTorr to 0.3 {\AA}/sec for $P_{Ar}$ = 8.5 mTorr, measured by a crystal monitor. The Crystal structure of deposited W films (22 nm) under different $P_{Ar}$ were characterized by Bruker x-ray diffraction spectroscope (XRD) with a resolution of $2^{0}$ as shown in Figure \ref{XRD}. $\rho_{W}$ of W films exhibited an increase from 10 $\mu\Omega$-cm to 750 $\mu\Omega$-cm as $P_{Ar}$ was raised from 4.5 mTorr to 8.5 mTorr [Figure \ref{Grain}(b)]. The W films tend to be more amorphous as $P_{Ar}$ increases. W films grown at higher $P_{Ar}$ ($\rho_{W}$ ranging from 750 to 350 $\mu\Omega$-cm) have single phase $\beta$-W which is expected to be A15 type crystal structure \cite{hao2015beta,lee2016growth}. This phase was characterized by the distinct peaks at 2$\theta$ = $35.6^{0}$, $39.6^{0}$, $43.8^{0}$ corresponding to $\beta$ (200), (210), (211) planes, respectively. However, in case of the W films with $\rho_{W}$ = 200, 100 $\mu\Omega$-cm, there was a subtle indication of $\alpha$ phase, evidenced by a peak shift from 2$\theta$ = $39.6^{0}$ to 2$\theta$ = $40.1^{0}$, which is a signature of $\alpha$-W (110) peak. These films are dominated by $\beta$ phase mixed with a minor presence of $\alpha$ as shown in Figure \ref{XRD}(a). As the $P_{Ar}$ was further reduced ($P_{Ar}$ $\leq$ 5 mTorr), $\alpha$ phase gradually became dominant. The as-deposited W film ($\rho_{W}$ = 30 $\mu\Omega$-cm) showed a highly crystalline structure dominated by $\alpha$ phase corresponding to $\alpha$ (110), (200), (211) planes at $2\theta$ = $40.4^{0}$, $58.3^{0}$, $73.6^{0}$ respectively, mixed with weaker $\beta$ peaks along (200), (211) at $2\theta$ =  $35.6^{0}$, $43.8^{0}$ respectively, indicating a mixed ($\alpha$ + $\beta$) phase. The transition from mixed ($\alpha$ + $\beta$)-W to pure $\alpha$-W occurred when the as-deposited film ($\rho_{W}$=30 $\mu\Omega$-cm) underwent annealing at $280^{0}$ C for 10 minutes, employing a temperature ramp-up duration of 3 hours and followed by a natural cooling of 6 hours in the vacuum chamber. The absence of $\beta$ diffraction peaks in the annealed film confirmed the existence of a pure $\alpha$ phase which possesses a bcc crystal structure characterized by the diffraction peaks corresponding to $\alpha$ (110), (200), and (211) at $2\theta$ = $40.4^{0}$, $58.3^{0}$, $73.6^{0}$ [Figure \ref{XRD}(b)]. XRD data provides a rough estimation of the grain size of W-thin films of different phases. According to Scherrer's equation \cite{scherrer}, average grain size can be evaluated from the equation expressed as $g = K\lambda/\delta cos(\theta)$, where, g is the average grain size, K is a shape factor (approximately 0.9), $\lambda$ is the x-ray wavelength (1.54 ${\AA}$), $\delta$ is width of diffraction peak at half maxima and $\theta$ is Bragg's angle \cite{maqbool2005surface}. Analyzing our XRD data, we found that the grain size varied systematically with $P_{Ar}$, exhibiting a decrease as $P_{Ar}$ increased [Figure \ref{Grain}(a)]. For further characterization, we measured the resistivity of W-thin films using the van der Pauw technique. The resistivity $\rho_{W}$ increased with $P_{Ar}$ stabilized with a maximum value of $\rho$ = 750 $\mu\Omega$-cm at 8.5 mTorr, while maintaining a uniform thickness of 22 nm. It is reported that $\beta$-W films tend to have notably high resistance because of the A15 crystal structure which is associated with pronounced electron-phonon scattering \cite{pai2012spin,choi2011}. However, $\alpha$-W is recognized for its low resistivity, and in our study we obtained a minimum of $\rho_{W}$ = 10 $\mu\Omega$-cm which is identified as pure $\alpha$ phase.\\

\noindent \textbf{Characterization of Py-Thin Films (FM layer)}:\\

\begin{figure}[h!]
\centering
\includegraphics[width=0.8\textwidth]{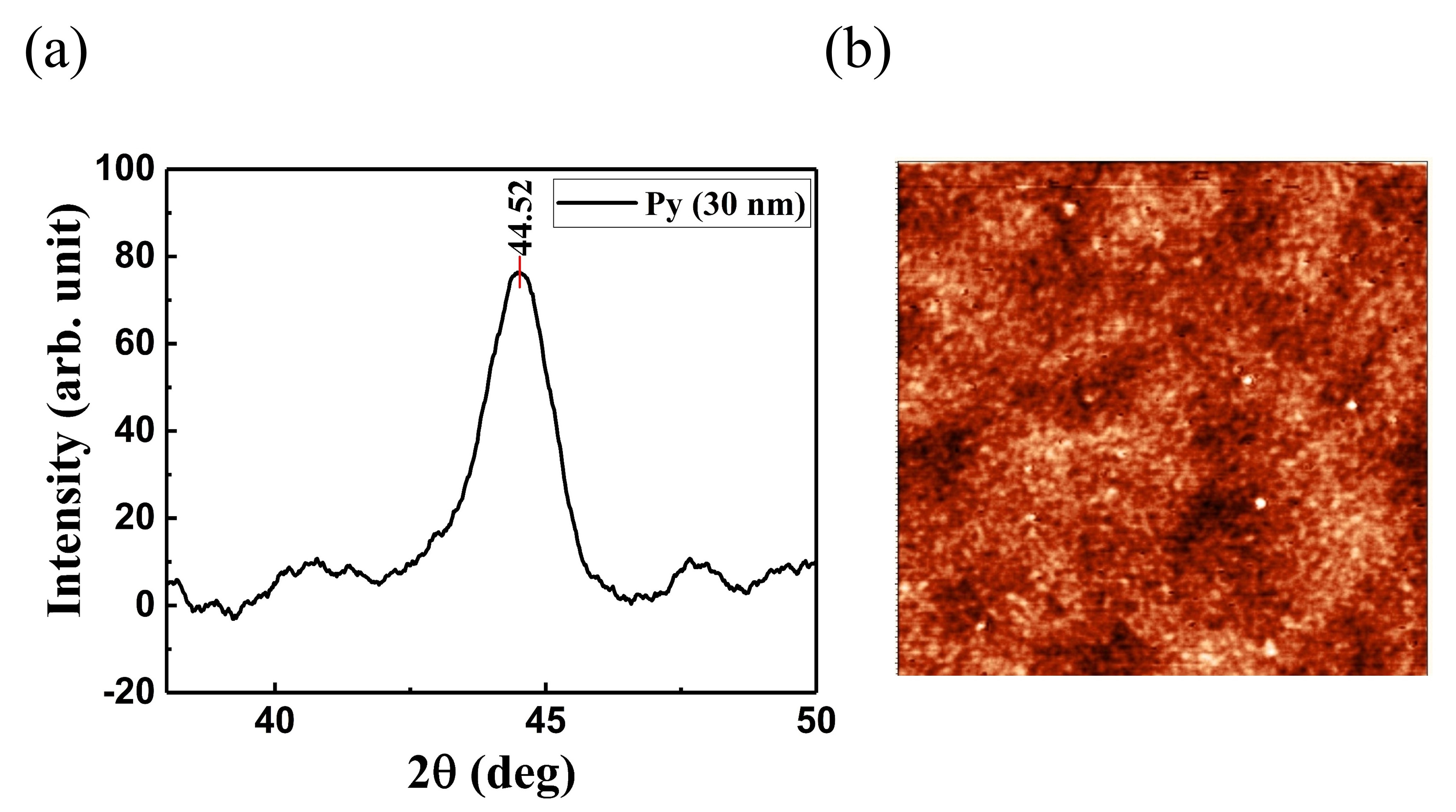}
\caption{(a) XRD image of Py thin film of thickness 30 nm. The observed peak at $2\theta$ = $44.52^{0}$ confirms the presence of fcc crystal structure, (b) AFM image captures the surface of Py thin film, covering a scanning area of 2 $\mu m$ $\times$ 2 $\mu m$.}
\label{Charaterization_Py}
\end{figure}

\noindent Py films were deposited on top of the HM layer utilizing the thermal evaporation technique. Prior to each deposition, an ion milling has been performed using dry argon plasma to clean the surface of the HM layer and to make the interface transparent. The specific parameters for the permalloy deposition are: $P_{base}$ $\approx$ $1\times 10^{-7}$ Torr, $P_{dep}$ $\approx$ $1\times 10^{-6}$ Torr, $T_{dep}$ $\approx$ $1300^{0}$C, deposition rate = (0.8-1.0) A/sec. These deposition parameters remained consistent across all devices. Characterization of the Py films was carried out using XRD and AFM techniques. In Figure \ref{Charaterization_Py}(a), the XRD image of the Py film is presented, with an observed peak at $2\theta$ = $44.52^{0}$, confirming the presence of an fcc crystal structure. Figure \ref{Charaterization_Py}(b) displays the AFM image depicting the surface morphology of the Py film over a scanning area of 2 $\mu m$ $\times$ 2 $\mu m$. The observed surface roughness was found to be 0.3 nm.\\

\noindent \textbf{Device Fabrication Procedure}:\\

\noindent A four-step deposition process including a contact pad was proposed for device fabrication, where each step involved photo-lithography (PL), RF ion etching, material deposition, and lift-off after deposition. (i) 
\textbf{Step 1: HM Channel:} A 300 $\mu m$ $\times$ 30 $\mu m$ rectangular bar was patterned  on $Si/SiO_{2}$ (300 nm) substrates by Photolithography, followed by RF Ar ion etching to create a trench in the `insulating' $SiO_2$ substrate. Etching parameters were $P_{Ar}$ = $6\times10^{-4}$ Torr, Power = 80 watt and ion current = 30 mA. Subsequently, HM materials (W and Pt) were deposited by dc magnetron sputtering, followed by lift off. The etch rates and deposition rates were calibrated to ensure that only the top surface of the HM channel is exposed for later steps. A series of W-thin films of thickness $22$ nm were used as HM channel material with a systematic variation of resistivity ranging from (10-750) $\mu\Omega$-cm by controlling by the $P_{Ar}$ during deposition, keeping sputtering power constant at 100 watt. A control device of Pt ($t_{Pt}$ = 37 nm, $\rho_{Pt}$ = 34 $\mu \Omega$-cm) HM channel was deposited by magnetron sputtering. The parameters used for Pt deposition were $P_{Ar}$ = $6\times10^{-3}$ Torr, power = 50 watt. (ii) \textbf{Step 2: Reference Metal(RM):}  Rectangular patterns of dimensions 20 $\mu m$ $\times$ 250 $\mu m$, aligned perpendicular to the long axis of HM and having overlap of $\sim 5-7 \mu m$. The RM material is essentially same as that of HM but with a constant low resistivity $\sim 60 \mu\Omega$-cm and thickness $t_{RM}$ = 37.5 nm. RM deposition is preceded by an ion milling was performed to clean the film surface and enhance interface transparency. (iii) \textbf{Step 3: Ferromagnetic Metal (FM)} Pattern dimensions are same as that of RM, and aligned linearly with the RM electrode but on the other side of the HM with overlap $\sim (5-7) \mu m$. Permalloy is chosen as the FM materials for all the devices, deposited by thermal evaporation from a RADAK furnace,  $P_{dep}$ = $1\times10^{-6}$ Torr, rate = 1.0 {\AA}/s. Same thickness and resistivity  of FM was  maintained for all devices , $t_{FM}$ = 30nm, $\rho_{FM}\sim  60\mu\Omega$-cm.  FM deposition was preceded by an Ar ion milling. The in plane and out of plane coercive fields of Py was found to be $\sim 5$mT and $\sim 0.8T$ respective and the hysteresis curves of AMR and PHE confirms in-plane magnetisation anisotropy, typical of Py films. (iv) \textbf{Step 4: Contact pads}  Cr-Au contact pads was deposited using electron-beam evaporation, overlapping the two ends of the HM channel and on free ends of RM and FM electrodes. All the resistivity vales are obtained by Van der Pauw method on a separate square piece at room temperature.\\

\noindent \textbf{Measurements}

\begin{figure}[h!]
\centering
\includegraphics[width=0.7\textwidth]{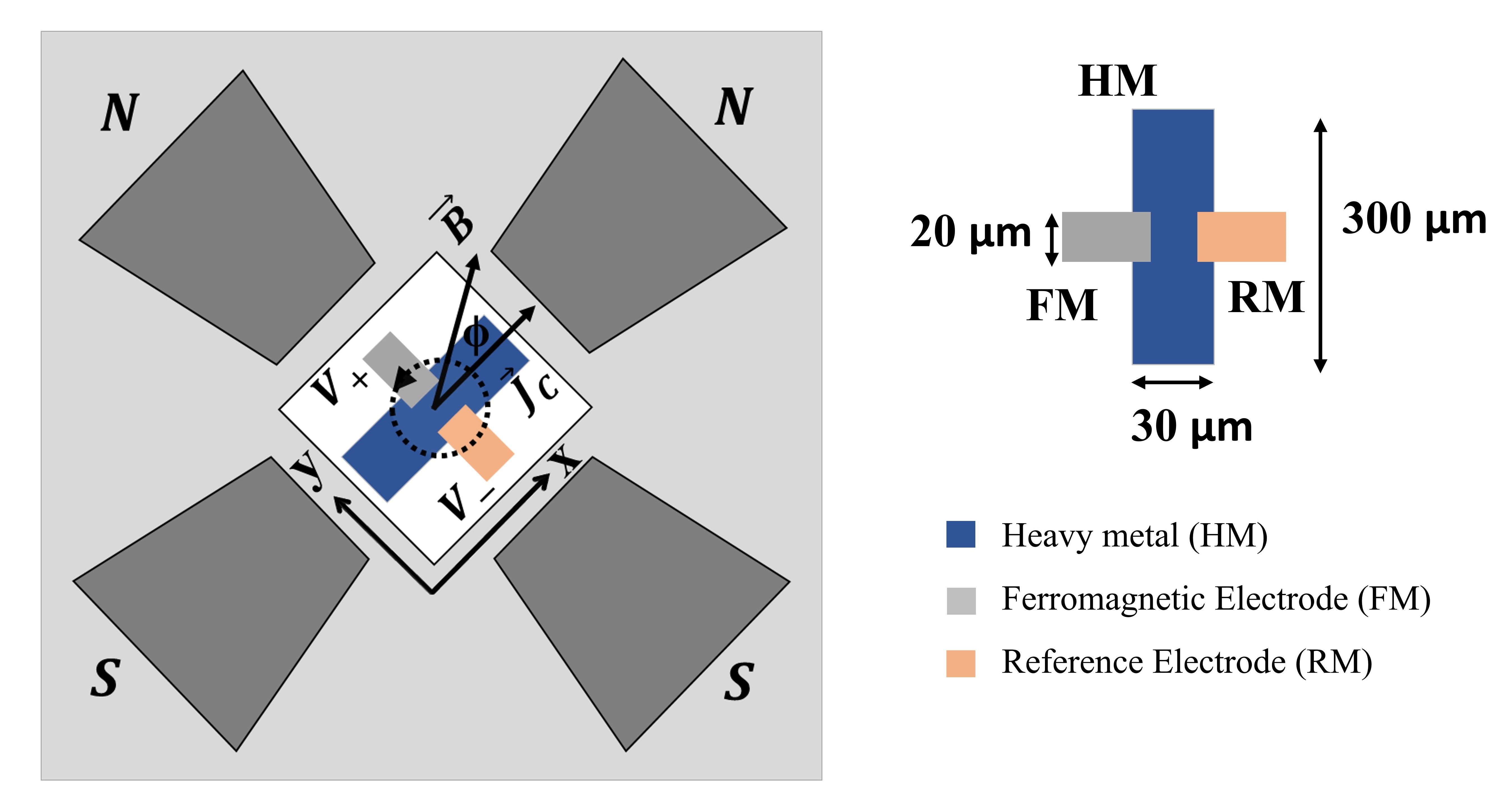}
\caption{Schematic presentation of a spin Hall device loaded inside a vector magnet.}
\label{Device_vector_magnet}
\end{figure}

\noindent Transport measurements were performed at room temperature using a four pole vector electromagnet magnet (Dexing, China) (Figure \ref{Device_vector_magnet}). The magnetic field values were calibrated separately for each pole pair with a DTM-151 Digital Teslameter and the field variation achieved for each pole were between $\pm0.5T$ with an accuracy of 0.1mT. The hysteresis scans were limited upto $\pm0.4T$ and step size of field variation was 1mT. 
DC current was applied using a keithley 6221 AC-DC  current source and  voltages measured using a 2182A Nano-voltmeter. Each resistance reading was performed with currents of opposite polarity and averaging the values obtained 20 times. Typical current density achieved in the devices are found in the range of $(0.5 - 7.0) \times 10^{9}$ $A/m^2$ that limits the joule heating power to $\sim$ 0.5 mW.

\clearpage

\noindent \textbf{Note 2. Charge Current Distribution Due to Shunting in FM and RM Electrodes}:\\

\noindent In our measurement, we obtain the $R_{xy}$ from the ratio $V_{xy}$/$I$, where, $I$ represents the total charge current. Notably, the FM and RM electrodes only partially overlap the HM layer, meaning that only a portion of the total current passes through the HM layer, contributing to the spin Hall resistance. As both the FM and RM electrodes form Ohmic contacts with the HM channel, we must account for the shunting effect in these devices. The resistivity of both FM and RM electrodes ($\rho_{FM/RM}$ = 60 $\mu \Omega$-cm) is comparable to or much lower than that of the resistivity of HM channel ($\rho_{HM}$ = 10-750 $\mu\Omega$-cm). Therefore, a significant portion of the current is diverted through the FM and RM electrodes when it passes through the HM channel in the region where FM and RM are present. This shunting of current through the FM and RM electrodes reduces the current flowing through the HM channel.

\noindent We calculate the charge current along HM layer, considering the SHE configuration. However, same calculation is applied to ISHE. The relationship between the electric field ($\vec{E}$) and the current density ($\vec{J}_{C}$) is expressed as. 
\begin{equation}\label{eq:1} 
\vec{E}=\rho\vec{J}_{C}
\end{equation}
In our device, the electric field at the junctions HM/FM and HM/RM should be equal. Therefore, the current density through FM and RM can be rewritten using the equation \ref{eq:1}. 
\begin{eqnarray}\label{eq:2} 
\rho_{HM}J_{HM} = \rho_{FM}J_{FM}: J_{FM} = J_{HM}\frac{\rho_{HM}}{\rho_{FM}} \\
\rho_{HM}J_{HM} = \rho_{RM}J_{RM} : J_{RM} = J_{HM}\frac{\rho_{HM}}{\rho_{RM}}
\end{eqnarray}
In the SHE configuration, when a current $I$ is applied along the HM channel, it is distributed across the FM, HM, and RM as follows: 
\begin{multline}\label{eq:3}
\\I = I_{HM} + I_{FM} + I_{RM}\\
I= J_{HM}A_{HM} + J_{FM}A_{FM} + J_{RM}A_{RM}\\
I= J_{HM} (A_{HM} + \frac{\rho_{HM}}{\rho_{FM}}A_{FM} + \frac{\rho_{HM}}{\rho_{RM}}A_{RM})\\
\end{multline} 
Where, $A_{HM}$ = $w_{HM}t_{HM}$, $A_{FM}$ = $w_{FM}t_{FM}$, $A_{RM}$ = $w_{RM}t_{RM}$. $t_{HM/FM/RM}$ are thicknesses of HM, FM, RM layers, respectively. $w_{HM}$ is the width of HM layer, and $w_{FM}$, $w_{RM}$ are overlapped width of FM, RM layers on HM layer. The current density in HM layer is expressed as.
\begin{eqnarray}\label{eq:4}
J_{HM}  = \frac{I}{A_{HM}(1 + \frac{\rho_{HM}}{\rho_{FM}}\frac{A_{FM}}{A_{HM}} + \frac{\rho_{HM}}{\rho_{RM}}\frac{A_{RM}}{A_{HM}})}  
\end{eqnarray}
The current flowing the HM layer is:
  \begin{equation}\label{eq:5}
I_{HM} = \frac{I}{[1 + \frac{\rho_{HM}}{t_{HM}w_{HM}} (\frac{t_{FM}w_{FM}}{\rho_{FM}} + \frac{t_{RM}w_{RM}}{\rho_{RM}})]}
\end{equation}\\

\clearpage

\noindent \textbf{Note 3. Angular Dependence of all W-Devices in Inverse Spin Hall Effect Configuration}:\\
\begin{figure}[h!]
\centering
\includegraphics[width=0.8\textwidth]{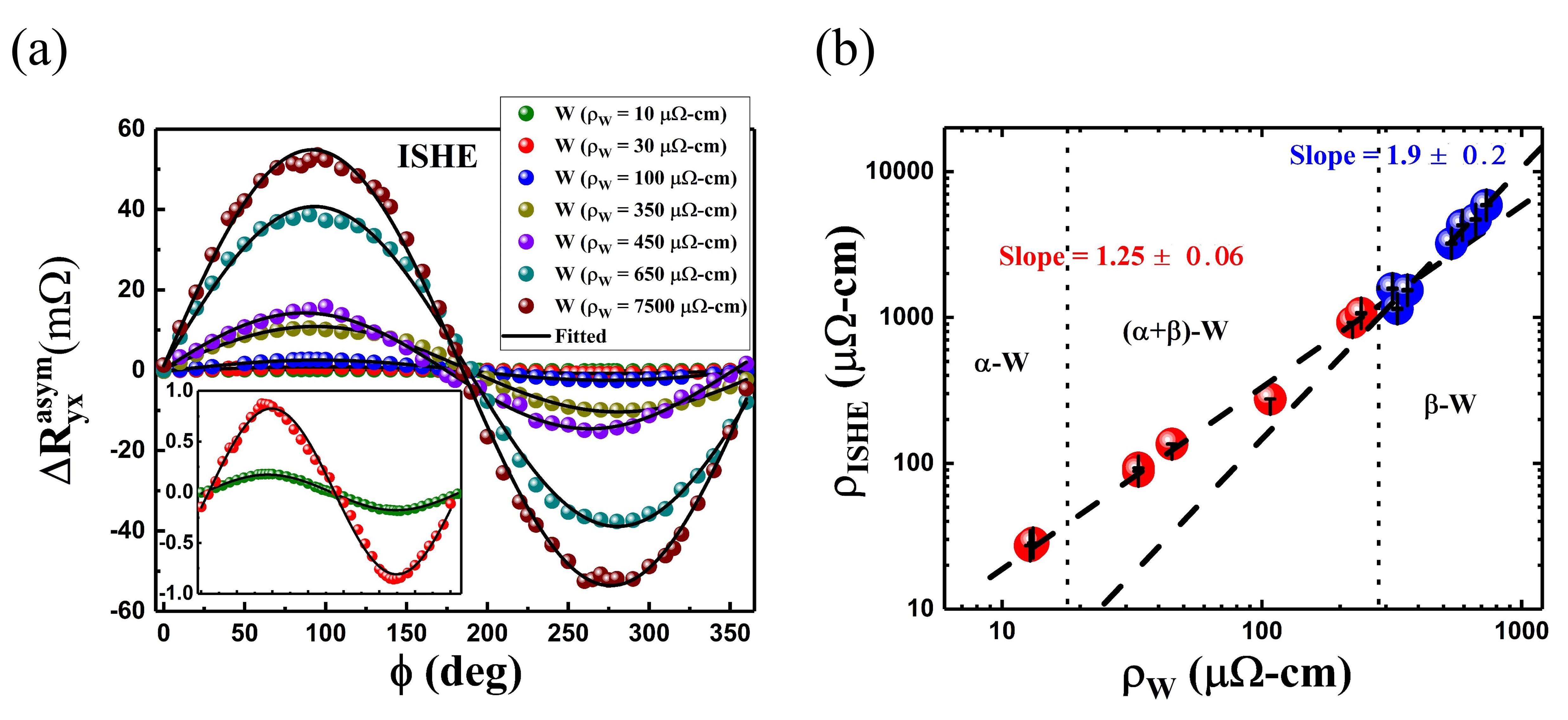}
\caption{(a) Shows the plots of $\Delta R^{asym}_{xy}$ as a function of angle $\phi$ in ISHE configuration for devices with $\rho_{W}$ = 10, 30, 100, 350, 450, 650, 750 $\mu\Omega$-cm. Each graph represents the result of an individual field scan of $R^{asym}_{xy}$ for a constant $\phi$ range spanning from $0^\circ$ to $360^\circ$ in $10^\circ$ increments. To enhance clarity, the inset illustrates the low-resistivity devices plotted separately. The solid lines depict fits to the equation $R_{SHE} \sin(\phi-\phi_{0})$, (b) The spin Hall resistivity ($\rho_{ISHE}$) is plotted against the longitudinal resistivity ($\rho_{W}$) on a logarithmic scale for all devices. Pure $\beta$-W data points are highlighted in blue, while those for pure $\alpha$-W and mixed ($\alpha+\beta$)-W are shown in red. Each data set is individually fitted to straight lines, with the corresponding slopes indicating the power law exponents.}
\label{Scaling}
\end{figure}

The angular variation of $\Delta R^{asym}_{yx}$ with different W resistivity of $\rho_W$ = 10, 30, 100, 350, 450, 650, 750 $\mu\Omega$-cm is plotted in Figure \ref{Scaling}(a). The inset graph shows the plots for devices with $\rho_W$ = 10 and 30 $\mu\Omega$-cm, as the magnitudes of $\Delta R_{yx}^{asym}$ is comparatively small and distinguishable in the main plot. Similar to SHE, variation of $\Delta R_{yx}^{asym}$ is fitted with curves of the form $R_{SHE} sin(\phi-\phi_{0})$. Figure \ref{Scaling}(b) shows the plot of $\rho_{ISHE}$ as a function of $\rho_{W}$ on log-log scale. The slope extracted for all W-devices is found to be 1.25 $\pm$ 0.04. When they are plotted separately, the slopes for pure $\beta$-W and the combined $\alpha$-W, mixed ($\alpha+\beta$)-W are 1.25 $\pm$ 0.06 and 1.9 $\pm$ 0.2, respectively.


\end{document}